
\input phyzzx
\date={1993}
\rightline {April 1993} \rightline {SUSX--TH--93/12.}
\title {Yukawa Couplings involving Excited Twisted Sector States
for ${\bf Z}_N$ and ${\bf Z}_M\times {\bf Z}_N$  Orbifolds.}
\author{D. Bailin$^{a}$, \ A. Love$^{b}$ \ and \ W.A. Sabra$^{b}$}
\address {$^{a}$School of Mathematical and Physical
Sciences,\break
University of Sussex, \break Brighton U.K.}
\address {$^{b}$Department of Physics,\break
Royal Holloway and Bedford New College,\break
University of London,\break
Egham, Surrey, U.K.}
\abstract{We study the Yukawa couplings among excited twist fields
which might arise in the low-energy effective field theory
obtained by compactifying the heterotic string on ${\bf Z}_N$ and
${\bf Z}_M\times {\bf Z}_N$ orbifolds.}
\endpage
\REF\one{L. Dixon, J. A. Harvey, C. Vafa and E. Witten, Nucl. Phys.
B261 (1985) 678; B274 (1986) 285.}
\REF\font {A. Font, L.E. Ibanez and F. Quevedo, Phys. Lett B217 (1989) 272.}
\REF\two{ A. Font, L. E. Ibanez,
F. Quevedo and A. Sierra, Nucl. Phys. B331 (1991) 421.}
\REF\japanese{Y. Katsuki, Y.Kawamura, T. Kobayashi and N. Ohtsubo, Phys. Lett.
B212 (1988) 339;
Y. Katsuki, Y.Kawamura, T. Kobayashi, Y. Ono, K. Tanoika and N. Ohtsubo,
Phys. Lett.
B218 (1989) 444.}
\REF\three{L. Dixon, D. Friedan, E. Martinec and S. Shenker, Nucl.
Phys. B282 (1987) 13.}
\REF\four{S. Hamidi, and C. Vafa, Nucl. Phys. B279 (1987) 465.}
\REF\five{ L. E. Ibanez, Phys. Lett. B181 (1986) 269.}
\REF\six{J. A. Casas and C. Munoz, Nucl. Phys. B332 (1990) 189}
\REF\seven {J. A. Casas, F. Gomez and C. Munoz,  Phys. Lett.
B251(1990) 99}
\REF\ja{Y. Katsuki, Y.Kawamura, T. Kobayashi, N. Ohtsubo, Y.Ono and
K. Tanioka, Nucl. Phys. B341 (1990) 611; D. Narkushevich, N. Olshanetsky
and A.Perelonov, Comm. Math. Phys. 111 (1987) 247.}
\REF\eight {J. A. Casas, F. Gomez and C. Munoz,  CERN preprint,
TH6194/91.}
\REF\nine{T. Kobayashi and N. Ohtsubu, Kanazawa
preprint, DPKU--9103.}
\REF\ten {T. T. Burwick, R. K. Kaiser and H. F.
Muller, Nucl. Phys. B355 (1991) 689}
\REF\eleven{J. Erler, D. Jungnickel and J. Lauer, Phys. Rev D45
(1992) 3651; S. Stieberger, D. Jungnickel, J. Lauer
and M. Spalinski, preprint, MPI--Ph/92--24; J. Erler,  D.
Jungnickel, M. Spalinski and S. Stieberger, preprint,
MPI--Ph/92--56.}
\REF\twelve{D. Bailin, A. Love and W. A. Sabra,
Mod. Phys. Lett A6 (1992) 3607; D. Bailin, A. Love and W. A. Sabra,
Sussex preprint, SUSX-TH-92/17, to be published in Nucl. Phys. B.}
\REF\thirteen{S. Stieberger, preprint, TUM--TH--151/92.}
\REF\David {D. Bailin, A. Love and W. A. Sabra, SUSX-TH-93/11, to be
published in Phys. Lett. B.}
\REF\fourteen{V. S. Kaplunovsky, Nucl. Phys. B307 (1988) 145; L. J.
Dixon, V. S. Kaplunovsky and J. Louis,  Nucl. Phys. B355 (1991) 649;
J. P. Derenddinger, S. Ferrara, C. Kounas and F. Zwirner, Nucl.
Phys. B372 (1992) 145, Phys. Lett. B271 (1991) 307.}
\REF\fifteen{I. Antoniadis, J. Ellis, R. Lacaze and D. V.
Nanopoulos, Phys. Lett. B268 (1991); S. Kalara, J. L. Lopez
and D. V. Nanopoulos,  Phys. Lett. B269 (1991) 84.}
\REF\seventeen{L. E. Ibanez and D.
L\"ust, Nucl. Phys. B382 (1992) 305.}
\REF\jap{T. Kobayashi and N. Ohtsubo, Phys. Lett. B262 (1991) 425.}
\REF\nineteen{A. Font, L. E.
Ibanez, H-P. Nilles  and F. Quevedo,  Nucl. Phys. B307 (1988)
109.}
\REF\realistic{L. E.
Ibanez, J. E. Kim,  H-P. Nilles and F. Quevedo,  Phys. Lett. B191 (1987)
3.}
\REF\twenty{A.A. Belavin, A.M. Polyakov and A.B. Zamolodchikov,
Nucl. Phys. B241 (1984) 333.}
\chapter {Introduction}
Of all known conformal field theories, orbifold models [\one,\two]
offer the best solution to the heterotic string equations of motion
with regard to a phenomenologically promising string
background.
The two-dimensional conformal orbifold field theory is
described by the Euclidean action
$$S={1\over\pi}\int d^2z
\Big(\partial_{z}{X^i}\partial_{\bar z}{\bar
X^i}+\partial_{\bar z}{X^i}\partial_{z}{\bar
X^i}\Big),\eqn\action$$
with $X^i$ the three complexified coordinates of the orbifold
target space $\cal O$. This space can be described as a six
dimensional Euclidean space $\cal E$ with its points identified under
the action of a space group $\cal S$ whose elements consists of both
discrete rotations and translations. These elements
are represented by the pair $(\alpha, l)$, where $\alpha$
constitutes the point group of the orbifold and $l$ is a
vector taking values on a lattice on which the point group acts as
an automorphism.
The generator of the point group, denoted by $\theta,$ must be a discrete
subgroup of $SU(3)$
in order to obtain an effective four dimensional theory with $N=1$
supersymmetry. If one restricts attention to abelian groups,
two possibilities arise: ${\bf Z}_N$ and
${\bf Z}_M\times {\bf Z}_N$. A ${\bf Z}_N$ group is generated by a
twist $\theta$ with $\theta^N=1$
and the point group is then given by $P=\{\alpha=\theta^n; n=0,1,..., N-1\}.$
The action of $P$ on the complex coordinates $X^i$ is defined by
$$\theta X^i=e^{2\pi iv^i}X^i, \qquad \sum_{i}v^i=0.\eqn\jeeves$$
Similarly ${\bf Z}_M\times {\bf Z}_N$ is  generated by two twists $\theta$ and
$\omega,$
with $\theta^M=1$, $\omega^N=1$ where $N$ is a multiple of $M$. Details of
the study and classification of these orbifolds can be found in
[\one, \two, \three].
For a particular model, it is always possible to find at least
one lattice on which the point group acts
as an automorphism. In the case of ${\bf Z}_M\times {\bf Z}_N,$ both point
groups should be realized as automorphisms on  the same lattice.

In addition to the twists acting on the left-moving supersymmetric sector
of the heterotic string, a gauge twisting
group ${\cal G}$ is also introduced.
This group acts on the right-moving gauge degrees of freedom
which, in the bosonic representation, can be realized in terms of shifts $V$ on
the $E_8\times E_8$ lattice. These shifts are restricted by the modular
invariance of
the orbifold partition function [\one,\japanese].
For standard embedding of the point group into $\cal G$ one obtains
models with left and right global world sheet N=2 supersymmetry -- $(2,2)$
orbifolds.
A generic choice of embedding (with or without Wilson lines) results in
models with only left-moving N=2 supersymmetry -- $(0,2)$ orbifolds.
Models with Wilson lines break the gauge group to a lower rank giving
semi-realistic four dimensional models [\realistic].

The identification of the point $x$ of the orbifold
with $\alpha x+l$ implies that a twisted string in the $\alpha$ twisted
sector is closed due to the action of the element $g=(\alpha, (1-\alpha)f)$,
where $f$ is a fixed point or torus under the action of the point
group element
$\alpha$. More precisely,
$g$ represents a conjugation class $hgh^{-1}$, $h\in {\cal S}$ which for prime
orbifolds
takes the form $(\alpha, (1-\alpha)(f+\lambda))$, where $\lambda$ is an
arbitrary
lattice vector.
For every fixed point (or torus) $f$ in a given twisted sector $\alpha$
one associates a vacuum state created by the action of a twist field
$\sigma_{\alpha,f}$ on the $SL(2,C)$ Neveu-Schwarz (NS) invariant vacuum.
These fields are analogous to the spin fields which transform  an NS vacuum
into a Ramond (R) vacuum. However, unlike spin fields, twist fields have no
realizations in terms of free fields
and are normally defined in terms of their operator product expansions
with the complex string coordinates associated with the orbifold
target space [\three].

In the various twisted sectors, massless states are created by
vertex operators constructed from the heterotic string degrees of freedom and
the
twist fields $\sigma_\alpha$. In general these states are not physical
and one has to take a linear combination of the twist operators.
Also, one could construct massless states with excitations.
The associated vertex operators will then contain, in addition to the
other string degrees of freedom, the so-called excited twist fields
defined by the operator product expansions
of ground state twist operators with $\partial_z X^i$ and
$\partial_z{\bar X}^i$.
The allowed number of oscillators present in the massless excited
states in a given sector can be determined by using the mass formula
for right movers
$${m^2\over 8}=N_{osc}+h_{KM}+h_{\sigma_\alpha}-1,\eqn\owen$$
where $N_{osc}$ is the fractional oscillator number, $h_{\sigma_\alpha}$ is
the conformal dimension of the twist field $\sigma_{\alpha}$
and $h_{KM}$ is the contribution from the gauge part to the conformal
dimension of the matter fields.
In the general case, for a particular choice of embedding and Wilson lines,
one can write down an expression for
$h_{KM}$ in terms of the Casimirs and levels of the relevant gauge
Kac--Moody algebras involved.

A partial comparison of orbifold string compactified theories with
low energy physics requires the determination of their Yukawa couplings.
These Yukawa couplings could then be
employed to make connections with phenomenological studies such as the
determination of quark and lepton masses and the mixing angles.
Because no momenta in the twisted directions are allowed, these Yukawa
couplings split into two factors,
one determined by the string degrees of freedom and the other is purely a
twist correlator.
Of particular importance is the exponential dependence of the twist
couplings on moduli
[\three-\four]
because of its possible bearing on hierarchies [\five].

So far, the discussion in the literature has been mostly limited to the
couplings involving only twisted sector ground states, Although it
has been pointed out how excited twist field correlations can be
computed [\three]. Also, an example of such a calculation has been
given in [\four], though using different approach to that of [\three].
In this paper we shall extend the discussion to Yukawa couplings
involving twisted sector excited states that might arise in
${\bf Z}_N$ and ${\bf Z}_M\times {\bf Z}_N$ orbifolds.
We have already given a brief discussion for ${\bf Z}_M\times {\bf Z}_N$
orbifolds elsewhere [\David].
For phenomenological studies, massless particles with
the conformal dimensions of
the standard model particles will be relevant.
Therefore, we focus on massless states in the twisted
sectors with the $SU(3)\times SU(2)\times U(1)$
quantum numbers of quarks, leptons and electroweak Higgses.

It is of particular importance to be able to
include these excited states in view of the fact that string
threshold loop corrections to gauge coupling constants
[\fourteen,\fifteen], consistent with their low energy values, have so
far always involved modular weights for quarks and leptons requiring
the use of excited twisted sector states. The ${\bf Z}_N$ and ${\bf
Z}_M\times {\bf Z}_N$ orbifolds considered are those with the point
group  realised in terms of the
Coxeter elements (product of Weyl reflections) of Lie algebra
root lattices [\ja, \jap].
This work is organized as follows. Section 2 contains a brief review of
the basic features
of the conformal field theory of the twist fields.
Sections 3 and 4 deal with the classification
and the calculation of the Yukawa couplings involving excited twisted fields
in ${\bf Z}_N$ and ${\bf Z}_M\times {\bf Z}_N$  orbifolds respectively.

\chapter{The Conformal Field Theory of Orbifolds}
As mentioned in the introduction, a different vacuum state belongs to each
fixed
point or torus in a given twisted sector. These vacuum states are created by
twist fields $\sigma_\alpha$ defined in terms of the operator
product expansions
$$\eqalign{&\partial_z X \sigma_\alpha(w, \bar w)\sim
(z-w)^{-(1-{\eta}_\alpha)}\tau_\alpha(w,\bar
w),\cr
&\partial_{z}{\bar X}
\sigma_\alpha(w, \bar w)\sim
(z-w)^{-\eta_\alpha}\tau'_\alpha(w,\bar w),\cr
&\partial_{\bar z} X \sigma_\alpha(w, \bar w)\sim
(\bar z-\bar w)^{-{\eta}_\alpha}\tilde\tau_\alpha(w,\bar
w),\cr
&\partial_{\bar z}{\bar X}
\sigma_\alpha(w, \bar w)\sim
(\bar z-\bar w)^{-(1-\eta_\alpha)}\tilde\tau'_\alpha(w,\bar w),}\eqn\local$$
where $\alpha=e^{2i\pi\eta_\alpha}$ and $0<\eta_\alpha<1$.
The index $i$ labelling the $i$-th complex
plane of the 6-dimensional compact manifold has been suppressed in \local.
Clearly, \local\ also serves as a definition of four different types
of the so-called excited twist fields
\foot { Note that we are using the definitions
of the excited twist fields
$\tau_\alpha$, $\tau'_\alpha$, $\tilde\tau_\alpha$
and $\tilde\tau'_\alpha$ used in reference [\ten], rather than that used
in our previous paper [\David].}.
The fields $\sigma_\alpha$ are primary conformal fields satisfying the
operator product expansion
$$T(z)\sigma_\alpha(w)\sim {h_{\sigma}\sigma_{\alpha}(w)\over (z-w)^2}+
{\partial_w\sigma_{\alpha}(w)\over (z-w)}+...,\eqn\stress$$
where $T(z)$ is the right moving stress energy tensor of the underlying
conformal field
theory. A similar expression holds for the left moving stress energy
tensor $\bar T(\bar z).$
The conformal dimensions of the field $\sigma_\alpha$ are given by
$(h_{\sigma_\alpha},\bar h_{\sigma_\alpha})=
\Big({1\over2}\eta_\alpha(1-\eta_\alpha),{1\over2}\eta_\alpha(1-\eta_\alpha)
\Big)$.
{}From \local\ one can read off the various conformal dimensions of the excited
twist fields.
For example, the conformal dimensions of
$\tau_\alpha$ and $\tau'_\alpha$ are
$(h_{\sigma_\alpha}+\eta_\alpha,\bar h_{\sigma_\alpha})$ and
$(h_{\sigma_\alpha}+1-\eta_\alpha,\bar h_{\sigma_\alpha})$ respectively.

Yukawa couplings involving twisted sectors ground states are given
by three-point functions involving fermionic and bosonic degrees of freedom.
However, the cruical dependence on the deformation parameters or moduli and
the particular fixed points or tori of the internal space
is entirely contained in bosonic
twist fields correlation functions
[\three, \four] of the type
$${\cal Z}=\prod_{i=1}^3{\cal Z}_i,\eqn\migra$$
with
$${\cal Z}_i=<\sigma_{\alpha,f_1}^i(z_1, \bar z_1)
\sigma_{\beta,f_2}^i(z_2, \bar z_2)\sigma_{\gamma,f_3}^i(z_3,
\bar z_3)>.\eqn\migran$$
Here $i$ is the index labelling the $i$-$th$ complex plane, $\alpha,$ $\beta$
and $\gamma$ are the point group elements for the three twisted
sectors involved and $f_1$, $f_2$ and $f_3$ are the corresponding fixed
points or tori.
The allowed trilinear couplings \migran\ are those with conjugation
classes whose product contains the identity. This condition gives the so-called
point group and space group selection rules. More precisely,
let $(\alpha, l_1)$, $(\beta, l_2)$ and $(\gamma, l_3)$ be space
group elements associated with the $\alpha$, $\beta$ and
$\gamma$ twisted sectors, where
$$l_1=(I-\alpha)(f_\alpha+\lambda_1),\eqn\n$$
$$l_2=(I-\beta)(f_\beta+\lambda_2),\eqn\m$$
$$l_3=(I-\gamma)(f_\gamma+\lambda_3),\eqn\p$$
and $\lambda_i$, $i=1,2,3,$
denotes an arbitrary lattice vector. Then, provided that
the point group selection rule
$$\alpha\beta\gamma=I\eqn\po$$
is already satisfied, the full space group selection rule contains
the additional
condition
$$l_1 +l_2+ l_3\ \hbox{contains}\ 0.\eqn\sel$$
Also, the Yukawa couplings involving ground states obey a
further selection rule coming from the factors $e^{ia.H}$ in the
left-moving part of the vertices,
where $H$ are the bosonic fields describing
the NSR left-moving fermions and $a$ is an $SO(10)$ weight.
This is the conservation of the $H$-lattice momentum, which is analogous
to the conservation of ordinary momenta $k$ which comes from the factors
$e^{ik^\mu x^\mu}$, with $x^\mu$ the four uncompactified string coordinates.

To solve the conformal field theory of the orbifold model it remains to
calculate
the three-point functions of all of its primary fields. The tools for
calculating the correlation functions of
ground twist fields (for the simplest choice of twist fields) in
orbifold models were first developed in [\three].
Twisted sector Yukawa coupling have been investigated for the
orbifolds ${\bf Z}_3$ [\three-\six], ${\bf Z}_7$ [\seven],
${\bf Z}_N$ [\eight-\eleven] and
${\bf Z}_M\times {\bf Z}_N$ [\twelve-\thirteen].

Roughly speaking, the method of evaluating the correlation functions in
orbifold models [\three] is as follows. One splits the
coordinate fields $X^i$ into a classical and a quantum piece,
$$X^i(z,\bar z)=X^i_{cl}(z,\bar z)+X^i_{qu}(z,\bar z),\eqn\max$$
where $X^i_{cl}$ is a classical solution of the
equation of motion and $X^i_{qu}$ are quantum fluctuations around the
instanton classical solutions.
The fact that the action \action\ is bilinear
in the fields $X^i$ allows the splitting of the  path integral representation
of
a four-point function into two factors, one
representing the instanton solution, and the other describing
the quantum fluctuations. This is expressed as
$${\cal Z}(x,\bar x) ={\cal Z}_{qu}(x,\bar x)\sum_{X_{cl}}e^{-S
(X_{cl})},\eqn\leo$$
where $x$ is the location of one of the vertices on the world sheet not fixed
by $SL(2,C)$ invariance.

Transporting the field $X^i$ around a closed path ${\cal C}$,
encircling twist fields with net zero twist, the following global
monodromy conditions are obtained,
$$\eqalign{&\Delta_{\cal C}X^i_{qu}=\oint_{\cal C}dz\partial_z X^i_{qu}+
\oint_{\cal C}d\bar z\partial_{\bar z} X^i_{qu}=0,\cr
&\Delta_{\cal C}X^i_{cl}=v^i,}\eqn\monodromy$$
where the complex vectors $v^i$ belong to the lattice coset which
is obtained by multiplying the space group elements corresponding
to twist vertices encircled by the contour ${\cal C}$.

Using the techniques of conformal field theory [\twenty] together with the
local and
global monodromy conditions (eqns \local\ and \monodromy),
the four point function involving four ground twist fields
is exactly solved. The operator product
coefficients, i.e., the
Yukawa couplings, are then deduced via the
appropriate factorization of the four point function.
\vfill\eject
\chapter{Excited Twisted Sector Yukawa couplings for
${\bf Z}_M\times {\bf Z}_N$ Orbifolds}
In this section the calculation of the Yukawa couplings involving
excited twisted states in the case of ${\bf Z}_M\times {\bf Z}_N$ orbifolds
is considered.
{}From the phenomenological point of view, only the conformal dimensions of
the standard model particles are relevant.
Therefore, massless states in the twisted
sectors with the $SU(3)\times SU(2)\times U(1)$
quantum numbers of quarks, leptons and electroweak Higgses are considered.
Then from \owen, the following restrictions on the allowed
excitations in a given twisted sector are obtained [\seventeen],
$$\eqalign{& N_{osc}\le 1 - h_{\sigma_\alpha}-{3\over5},\qquad {\hbox{for}}\ Q,
\ u_c\
{\hbox {and}}\ e_c\cr & N_{osc}\le 1 -h_{\sigma_\alpha}-{2\over5},\qquad
{\hbox{for}}\ L,\ d_c\
{\hbox{and}}\ H,}\eqn\mig$$
(where we have chosen the Kac-Moody levels, $3/5 k_1=k_2=k_3=1$.)
The inequality in \mig\ reflects the generic occurrence in orbifold models,
before spontaneous symmetry breaking, of extra $U(1)$ gauge fields to
which the quarks and leptons couple [\realistic].
The above arguments guarantee the
masslessness of the states. However
whether or not they are present in the massless spectrum of the theory is
determined by the generalized GSO projection of the orbifold
symmetries [\one, \font, \two].

The non-zero Yukawa couplings involving ground states obey a point group
selection rule and a selection rule for the $H$-lattice momentum associated
with bosonized NSR
fermion degrees of freedom. These selection rules have already
been written down in [\jap].
Allowed Yukawa couplings involving excited twisted sectors can be obtained from
those of ground states with the insertions of $\partial_z X$ and
$\partial_{z} \bar X$ vertices. The number of such insertions is restricted
by the
discrete symmetries of the two dimensional
sub-lattices of the six dimensional compact manifold [\four, \nineteen].
This means that assuming the discrete symmetry
acting on the $i$-$th$ complex plane is of order $N$, the correlation functions
involving
$(\partial_zX^i)^m(\partial_z \bar X^i)^n$ are allowed only if
$$m-n=0\ {\hbox {mod}}\ N.\eqn\selection$$
As an illustration, consider the case of ${\bf Z}_2\times {\bf Z}_6$ with
lattice
$SO(4)\times G_2^2.$ In the orthonormal basis, the twist's action
on the internal complex string degrees of freedom is given by
$$\eqalign{&\theta X^1=-X^1,\qquad \omega X^1=X^1,\cr
&\theta X^2=X^2,\qquad \omega X^2= e^{2\pi i/6}X^2,\cr
&\theta X^3=-X^3,\qquad \omega X^3= e^{-{2\pi i/6}}X^3,}\eqn\point$$
where $\theta$ and $\omega$ the generators of the point groups associated with
${\bf Z}_2$ and ${\bf Z}_6$ respectively. The twisted sectors in the theory
are $\omega$,
$\omega^2$,$\omega^3$, $\omega^4$, $\omega^5$, $\theta$, $\theta\omega$,
$\theta\omega^2$, $\theta\omega^3$.

Here the three-point functions among twisted sectors allowed by $SO(10)$
($H$-lattice) momentum conservation and point group selection rules
are given by
$$\eqalign{&\langle\sigma_\omega\sigma_{\theta\omega^2}
\sigma_{\theta\omega^3}\rangle,\
\langle\sigma_{\omega^2}\sigma_{\theta\omega}\sigma_{\theta\omega^3}\rangle,\
\langle\sigma_{\omega^3}\sigma_{\theta}\sigma_{\theta\omega^3}\rangle\cr
&\langle\sigma_{\omega^4}\sigma_{\theta}\sigma_{\theta\omega^2}\rangle,\
\langle\sigma_{\omega^4}\sigma_{\theta\omega}\sigma_{\theta\omega}\rangle,\
\langle\sigma_{\omega^5}\sigma_{\theta}\sigma_{\theta\omega}\rangle.}\eqn\yukawa$$
Using equation \mig, the allowed excitations in various sectors can be
determined and we list them as follows:
$$\eqalign{&\omega:\quad \partial_z X^2,
\partial_z X^2\partial_z X^2,\partial_z\bar X^3,\partial_z\bar X^3
\partial_z \bar X^3,\partial_z X^2\partial_z \bar X^3,\cr
&\omega^2:\quad \partial_z X^2, \partial_z \bar X^3,\cr
&\omega^4:\quad \partial_z \bar X^2, \partial_z X^3,\cr
&\omega^5:\quad \partial_z X^3,
\partial_z X^3\partial_z X^3,\partial_z\bar X^2,\partial_z\bar X^2
\partial_z \bar X^2,\partial_z X^3\partial_z \bar X^2,\cr
&\theta\omega:\quad \partial_z X^2,\cr
&\theta\omega^2:\quad \partial_z X^3.}\eqn\excitations$$
It can be easily shown that two possible non-trivial
excited three-point
functions can be obtained from the coupling
$\langle\sigma_{\omega^4}\sigma_{\theta\omega}\sigma_{\theta\omega}\rangle$,
namely
$\langle{\tau'}^2_{\omega^4}{\tau}^2_{\theta\omega}
\sigma^2_{\theta\omega}\rangle$ and
$\langle{\tau'}^2_{\omega^4}\sigma^2_{\theta\omega}{\tau}^2_{\theta\omega}\rangle.$
The rest of the allowed couplings are simply two point functions.

Following the same procedure as that outlined above, it can be easily
demonstrated
that no
non-trivial three-point functions exist for the orbifolds
${\bf Z}_2\times {\bf Z}_2$,  ${\bf Z}_2\times {\bf Z}_4,$
${\bf Z}_4\times {\bf Z}_4$ and ${\bf Z}_3\times {\bf Z}_3.$
The possible three-point functions involving excited couplings of all
other ${\bf Z}_M\times {\bf Z}_N$ orbifolds
are tabulated in the Appendix, using the notation $T_{pq}$ to denote the
$\theta^p\omega^q$
 twisted sector.
The excited twist fields which are found to be relevant in these cases are
$\tau^i_\alpha$,
$\tau'^i_\alpha,$
$\hat\tau^i_\alpha$ and $\hat\tau'^i_\alpha$, where the last two
are, respectively, the excited
twist fields created by applying $\partial_z X^i \partial_z X^i$ and $
\partial_z \bar X^i\partial_z \bar X^i$ on the $\alpha$-twisted vacuum.
However it is found that all the couplings
involving $\hat\tau_\alpha$ and $\hat\tau'_\alpha$ are merely
two point functions.
Thus we are interested in the computation of correlation functions of
the form
$${\Big({\cal Z}_i^3\Big)}_{excited}= <\tau_{\alpha}^i(z_1,
\bar z_1) \tau'{}^i_{\beta}(z_2, \bar z_2)\sigma_{\gamma}^i(z_3,
\bar z_3)>,\eqn\medi$$  up to  permutations of $\alpha$, $\beta$ and
$\gamma$. In \medi\ and subsequent equations we suppress the dependence of the
various fields on the
fixed points or tori and assume that the fields are located
at those satisfying the space group selection rule.

Normalization of the excited twist fields is necessary in order to create
normalized states.
The twisted mode expansion for $X^i$ and $\bar X^i$ suggests that the
normalization factors for  the excited fields
$\tau^i_\alpha$ and $\tau'^i_\alpha$ are $1/ \sqrt{2\eta_\alpha}$ and
 $1/\sqrt{2(1-\eta_\alpha)}$ respectively.
This normalization could also be determined from the explicit
calculations of the two-point function
$$\langle\tau^i_{\alpha^{-1}}(z_1, \bar
z_1){\tau'}^i_{\alpha}(z_2, \bar z_2)\rangle.\eqn\tpf$$
The evaluation of \tpf\ follows from that of the Green function [\three]
$$g(z,w,z_i)={-{1\over2}}
{\langle\partial_{z}{X^i}\partial_{w}{\bar
X^i}\sigma_{\alpha^{-1}}(z_1, \bar z_1)\sigma_{\alpha}(z_2,
\bar z_2\rangle\over\langle\sigma_{\alpha^{-1}}(z_1, \bar
z_1)\sigma_{\alpha}(z_2, \bar z_2\rangle}.\eqn\green$$
$g$ is completely fixed by the local properties of the string coordinates
in the presence of twists, encoded in the operator product expansions \local,
and the fact that
$$g(z,w,z_i)\sim {1\over (z-w)^2}+\hbox{finite}, \quad z\rightarrow w.
\eqn\add$$
Explicitly
$$\eqalign{g(z,w,z_i)&\sim (z-
z_1)^{-\eta_\alpha},\qquad z\rightarrow z_1,\cr
&\sim (z- z_2)^{-(1-\eta_\alpha)},\qquad  z\rightarrow z_2,\cr
&\sim (w-z_1)^{-(1-\eta_\alpha)},\qquad w\rightarrow z_1,\cr
&\sim (w-z_2)^{-\eta_\alpha},\qquad w\rightarrow z_2.}\eqn\va$$
Using \add\ and \va\ we conclude that
$$\eqalign{g(z,w,z_i)=&(z-
z_1)^{-\eta_\alpha} (z- z_2)^{-(1-\eta_\alpha)}
(w-z_1)^{-(1-\eta_\alpha)}
(w-z_2)^{-\eta_\alpha}\cr
&{\eta_\alpha(z-z_1)(w-z_2)+{(1-\eta_\alpha)}(z-z_2)(w-z_1)\over
(z- w)^2}.}\eqn\mgreen$$
Then \tpf\ is simply
$$\eqalign{\langle\tau^i_{\alpha^{-1}}(z_1, \bar
z_1){\tau'}^i_{\alpha}(z_2, \bar
z_2)\rangle=&-2\langle\sigma_{\alpha^{-1}}(z_1, \bar
z_1)\sigma_{\alpha}(z_2, \bar z_2)\rangle\cr
&\lim_{z\rightarrow z_1\atop w\rightarrow z_2}(z-
z_1)^{-\eta_\alpha}(w-\bar z_2)^{-\eta_\alpha} g(
z,w)\cr &
=2(1-\eta_\alpha)(-1)^{\eta_\alpha-1}
(z_1-z_2)^{-2(1-\eta_\alpha)}|
z_1-z_2|^{-4h_{\sigma_\alpha}}.}\eqn\zahra$$
The factors of $(-1)^{\eta_\alpha-1}$, $(z_1-z_2)$ and
$(\bar z_1-\bar z_2)$ are to be expected because of $SL(2,C)$
invariance.  This confirms the normalization factors for
$\tau^i_\alpha$ and ${\tau'}^i_\alpha$ as suggested by the twisted mode
expansion.

Similar calculations hold for the two-point function
$$\langle\hat\tau^i_{\alpha^{-1}}(z_1, \bar
z_1)\hat{\tau'}^i_{\alpha}(z_2, \bar z_2)\rangle,\eqn\dtpf$$
which can be obtained from knowledge of the correlation function
$$\eqalign{g'(z,w,z',w',z_i)=&{{1\over4}}
{\langle\partial_{z}{X^i}\partial_{z'}{X^i}\partial_{w}{\bar
X^i}\partial_{w'}{\bar
X^i}\sigma_{\alpha^{-1}}(z_1, \bar z_1)\sigma_{\alpha}(z_2,
\bar z_2\rangle\over\langle\sigma_{\alpha^{-1}}(z_1, \bar
z_1)\sigma_{\alpha}(z_2, \bar z_2\rangle}\cr
&=g(z,w,z_i)g(z',w',z_i)+g(z,w',z_i)g(z',w,z_i),}\eqn\dgreen$$
as being simply
$$\eqalign{&\langle{\hat\tau}^i_{\alpha^{-1}}(z_1, \bar z_1)
{\hat\tau}'^i_{\alpha}(z_2, \bar z_2)\rangle=
4\langle\sigma_{\alpha^{-1}}(z_1, \bar z_1)\sigma_\alpha(z_2, \bar z_2)\rangle
\cr &\lim_{{z,z'\rightarrow z_1}\atop {w,w'\rightarrow z_2}}
(z-z_1)^{-\eta_\alpha}(z'-z_1)^{-\eta_\alpha}
(w-\bar z_2)^{-\eta_\alpha}
(w'-\bar z_2)^{-\eta_\alpha}g'(z,w,z',w',z_i)\cr &
=8(1-\eta_\alpha)^2(-1)^{2\eta_\alpha-2}
(z_1-z_2)^{-4(1-\eta_\alpha)}|
z_1-z_2|^{-4h_{\sigma_\alpha}}.}\eqn\safia$$
This implies that the normalization factors for
$\hat\tau^i_\alpha$ and $\hat{\tau'}^i_\alpha$ are
$1/2{\sqrt 2}\eta_\alpha$ and
${1/2{\sqrt 2}(1-\eta_\alpha)}$ respectively.

Having normalized the relavent fields, we turn to the calculation of
the three point-function \medi.
This correlator can be written, using the operator
products expansions \local, in terms of the internal right-moving complex
string coordinates
fields as  $$\eqalign{{\cal Z}^3_{excited}=&\lim_{z\rightarrow
z_1\atop w\rightarrow z_2} (z- z_1)^{1-\eta_\alpha}
(w-z_2)^{\eta_\beta}\cr &
\langle\partial_z X\partial_w\bar X
\sigma_{\alpha}(z_1, \bar z_1)\sigma_{\beta}(z_2, \bar
z_2)\sigma_{(\alpha\beta)^{-1}} (z_3, \bar z_3)\rangle,}\eqn\thpf$$
where the index $i$ labelling the $i$-th
complex plane has been suppressed in this and subsequent equations.
Separating $X$ into a classical part and a quantum part as in \max,
we obtain
$$\eqalign{&\langle\partial_{z}{X}\partial_{w}{\bar
X}\sigma_{\alpha}(z_1, \bar z_1)\sigma_{\beta}(z_2, \bar
z_2)\sigma_{(\alpha\beta)^{-1}}(z_3, \bar z_3)\rangle=\cr
&\sum_{X_{cl}}e^{-S_{cl}} {\langle\partial_{
z}{X}_{qu}\partial_{w}{\bar X}_{qu}\rangle}_{{3-twists}} +
\sum_{X_{cl}}e^{-S_{cl}} {\partial_{z}{X}}_{cl}\partial_{w}{\bar
X}_{cl}{\cal Z}_{qu}^3,}\eqn\h$$ where
$$\eqalign{{\langle\partial_{z}{X}_{qu}\partial_{w}{\bar
X}_{qu}\rangle}_{{3-twists}}=&\int {\cal D}X_{qu}e^{-S_{qu}}
\partial_{z}{X}_{qu}\partial_{w}{\bar X}_{qu}\cr
&\sigma_{\alpha}(z_1, \bar z_1)\sigma_{\beta}(z_2, \bar
z_2)\sigma_{(\alpha\beta)^{-1}}(z_3, \bar z_3),}\eqn\he$$ $${\cal
Z}_{qu}^3= \int {\cal D}X_{qu}e^{-S_{qu}} \sigma_{\alpha}(z_1,
\bar z_1)\sigma_{\beta}(z_2, \bar
z_2)\sigma_{(\alpha\beta)^{-1}}(z_3, \bar z_3).\eqn\hel$$
The classical fields have
derivatives of the form
$$\eqalign{&\partial_{z}X_{cl}=d (z-z_1)^{-(1-\eta_\alpha)}
(z-z_2)^{-(1-\eta_\beta)}
(z-z_3)^{-(\eta_\alpha+\eta_\beta)},\cr &
\partial_{w}{\bar X}_{cl}=a (w-z_1)^{-\eta_\alpha}
(w-z_2)^{-\eta_\beta}
(w-z_3)^{-(1-\eta_\alpha-\eta_\beta)},}\eqn\pl$$ which are consistent with
the operator product expansions \local.
The constant $a$ must
be chosen to be zero for an acceptable classical solution because
otherwise the classical action is divergent. Consequently, the
second term in \h\ vanishes, and the moduli dependence of
${\cal Z}^3_{excited}$, which is contained in
${\sum}_{X_{cl}}e^{-S_{cl}}$, is exactly the same as for the
three-point function with unexcited twist fields.
In order to determine the overall
normalization of the three-point function, which depends on the
twisted sectors involved, we consider the calculation of
the four-point function, $${\cal Z}^4_{excited}=
{1\over2\sqrt{\eta_\beta(1-\eta_\alpha)}}\langle{\sigma}_{\alpha^{-1}}(z_1)
{\tau'_\alpha}(z_2)\sigma_{\beta^{-1}}(z_3)\tau_{\beta}(z_4)\rangle,\eqn\pla$$
where ${1/2 \sqrt{\eta_\beta(1-\eta_\alpha)}}$ is a normalization
factor due to the fields $\tau_\beta$ and $\tau'_\alpha.$
Again with  the aid of the
operator product expansions \local\ this can be written as
$$\eqalign{{\cal Z}^4_{excited}=&
{1\over2\sqrt{\eta_\beta(1-\eta_\alpha)}}\lim_{z\rightarrow z_4\atop
w\rightarrow z_2} (z-z_4)^{1-\eta_\beta} (w-z_2)^{\eta_\alpha}\cr
&\langle\partial_{z}{X}\partial_{w}{\bar
X}\sigma_{\alpha^{-1}}(z_1,{\bar z}_1) {\sigma_\alpha}(z_2, {\bar
z}_2){\sigma}_{\beta^{-1}}(z_3, {\bar z}_3) \sigma_{\beta}(z_4,
{\bar z}_4)\rangle\cr &
={1\over2\sqrt{\eta_\beta(1-\eta_\alpha)}}
\lim_{z\rightarrow z_4\atop
w\rightarrow z_2} (z-z_4)^{1-\eta_\beta} (w-z_2)^{\eta_\alpha}
\cr &\sum_{X_{cl}}e^{-S_{cl}}
{\langle\partial_{z}{X}_{qu}\partial_{w}{\bar
X}_{qu}\rangle}_{{4-twists}} + \sum_{X_{cl}}e^{-S_{cl}}
{\partial_{z}{X}}_{cl}\partial_{w}{\bar
X}_{cl}{\cal Z}_{qu}^4,}\eqn\plan$$
where
$$\eqalign{&{\langle\partial_{z}{X}_{qu}\partial_{\bar w}{\bar
X}_{qu}\rangle}_{{4-twists}}=\cr
& \int {\cal D}X_{qu}e^{-S_{qu}}
\partial_{z}{X}_{qu}\partial_{w}{\bar
X}_{qu}
\sigma_{\alpha^{-1}}(z_1,{\bar z}_1)
{\sigma_\alpha}(z_2, {\bar z}_2){\sigma}_{\beta^{-1}}(z_3, {\bar
z}_3) \sigma_{\beta}(z_4, {\bar z}_4)}\eqn\plants$$
and
$${\cal Z}_{qu}^4=
\int {\cal D}X_{qu}e^{-S_{qu}}
\sigma_{\alpha^{-1}}(z_1,{\bar z}_1)
{\sigma_\alpha}(z_2, {\bar z}_2){\sigma}_{\beta^{-1}}(z_3, {\bar
z}_3) \sigma_{\beta}(z_4, {\bar z}_4).\eqn\le$$
The first term in \plan\ can be evaluated from the Green function in the
presence of
four twists,
$$h(z, w,z_i)=
-{1\over2}{\langle\partial_{
z}{X}_{qu}\partial_{w}{\bar
X}_{qu}\sigma_{\alpha^{-1}}(z_1, \bar z_1)\sigma_{\alpha}(z_2,
\bar z_2)\sigma_{\beta^{-1}}(z_3, \bar
z_3)\sigma_{\beta}(z_4, \bar z_4\rangle\over\langle
\sigma_{\alpha^{-1}}(z_1, \bar z_1)\sigma_{\alpha}(z_2,
\bar z_2)\sigma_{\beta^{-1}}(z_3, \bar
z_3)\sigma_{\beta}(z_4, \bar z_4)\rangle}.\eqn\reem$$
Using the operator product expansions \local, and the fact that
$$h(z, w,z_i)\sim {1\over (z-w)^2}+\hbox{finite},
\quad z\rightarrow w,\eqn\operator$$
\reem\ can be shown to take the form
$$\eqalign{h(z, w,z_i)
=&(z-z_1)^{-\eta_\alpha}(z-z_2)^{\eta_\alpha-1}(z-z_3)^{-\eta_\beta}
(z-z_4)^{\eta_\beta-1}\cr &(w-z_1)^{\eta_\alpha-1}
(w-z_2)^{-\eta_\alpha}(w-z_3)^{\eta_\beta-1}
(w-z_4)^{-\eta_\beta}
\cr &\Big({P(z,w)\over
(z-w)^2}+A(z_i)\Big),}\eqn\poly$$
where the coefficients of the polynomial
$$P(z,w)=\sum_{i,j=0}^2a_{ij}z^iw^j,\eqn\zak$$
are determined using \operator\ and the freedom of defining the
finite part $A$. These are given by
$$\eqalign{a_{00}=& z_1z_2z_3z_4,\cr
a_{01}=&(k-1)z_2z_3z_4-kz_1z_3z_4+(l-1)z_1z_2z_4-lz_1z_2z_3,\cr
a_{10}=&-kz_2z_3z_4+(k-1)z_1z_3z_4-lz_1z_2z_4+(l-1)z_1z_2z_3,\cr
a_{11}=&
z_1z_4+z_2z_3+z_1z_2+z_3z_4,\cr
a_{20}=&z_1z_3-{k+l\over2}z_1z_3+{l-k\over2}z_1z_4+{k-l\over2}
z_2z_3+{k+l\over2}z_2z_4,\cr
a_{02}=&z_2z_4+{k+l\over2}z_1z_3+{k-l\over2}z_1z_4-
{k-l\over2}z_2z_3
-{k+l\over2}z_2z_4,\cr
a_{12}= &-kz_1+(k-1)z_2-lz_3+(l-1)z_4,\cr
a_{21}=& (k-1)z_1-kz_2+(l-1)z_3-lz_4,\cr
a_{22}= &1.}\eqn\coe$$
Unlike the Green function in the presence of two twists, $h$ can not be
entirely determined from the
local monodromy conditions
and one has to fix the function $A(z_i)$ using the global monodromy conditions
\monodromy. The determination of $A$ can be found in ref [\ten].
Thus we can write
$$\eqalign{\lim_{z\rightarrow z_4\atop w\rightarrow
z_2} (z-z_4)^{1-\eta_\beta} (w-z_2)^{\eta_\alpha}
&h(z,w,z_i)=\cr &(-1)^{-\eta_\beta}
x^{\eta_\alpha}(x-1)^{\eta_\beta-1}\Big((x-1)\partial_x
logI\Big),}\eqn\hell$$
where we have used $SL(2,C)$ invariance to set
$$z_1=0,\quad z_2=x,\quad z_3=1,\quad z_4=\infty,$$
$$A(z_i)=-{1\over z_\infty} A(x,\bar x)=-{1\over z_\infty}\Big(
{\eta_\alpha-\eta_\beta\over2}x+x(1-x)log I(x,\bar x)\Big) \eqn\a$$
and
$$I (x,\bar x)=J_2\bar G_1(x){H}_2(1-x)+ J_1{G}_2(x){\bar H}_1(1-\bar
x),\eqn\engla$$ with
$$J_1={\Gamma({\eta_\alpha})\Gamma(1-{\eta_\beta})
\over\Gamma(1+{\eta_\alpha}-{\eta_\beta})},\qquad
J_2={\Gamma(1-{\eta_\alpha})\Gamma({\eta_\beta})
\over\Gamma(1+{\eta_\beta}-{\eta_\alpha})},\eqn\engla$$
$$G_1(x)=F({\eta_\alpha},1-{\eta_\beta};1;x),\qquad
G_2(x)=F(1-{\eta_\alpha},{\eta_\beta};1;x)\eqn\englan$$ and
$$H_1(x)=
F({\eta_\alpha},1-{\eta_\beta};1+{\eta_\alpha}-{\eta_\beta};x),\qquad
H_2(x)=
F(1-{\eta_\alpha},{\eta_\beta};1+{\eta_\beta}-{\eta_\alpha};x).
\eqn\england$$
Finally, the second term in \plan\ can be evaluated by writing,
in agreement with the local monodromy behaviour,
$$\partial_z X_{cl}=c z^{-\eta_\alpha}(z-x)^{\eta_\alpha-1}
(z-1)^{-\eta_\beta}(z-z_\infty)^{\eta_\beta-1}$$
$$\partial_w X_{cl}=d w^{\eta_\alpha-1}(w-x)^{-\eta_\alpha}
(w-1)^{\eta_\beta-1}(w-z_\infty)^{-\eta_\beta}$$
The normalization factors $c$ and $d$ are then derived in terms of
hypergeometric functions
using global monodromy conditions.
The expressions for $c$ and $d$ can be found in ref [\ten].

To factorize, we take
$x\rightarrow\infty$. In this limit, the leading term of
${\cal Z}^4_{excited}$ should factorize into a product of the Yukawa coupling
\thpf\ and the Yukawa coupling
of unexcited twisted matter fields.
{}From the behaviour of the hypergeometic functions as $x\rightarrow
\infty$, the second term in \plan\ gives no contribution -- this is consistent
with
the remarks made after eq \pl.
Therefore it remains to factorize the quantum part of the first term in \plan\
which
will give the overall normalization factor (the quantum part) of the Yukawa
coupling \thpf.
In the limit $x\rightarrow \infty$, the quantum piece of the first term in
\plan\ should read
$$\lim_{x,\bar x\rightarrow\infty}\Big({\cal
Z}^4_{excited}\Big)_{qu}=\lim_{x,\bar x\rightarrow
\infty}
\langle\tau_\beta(\infty)\tau'_\alpha(x)
\sigma_{{(\alpha\beta)}^{-1}}(0)\rangle
\langle\sigma_{\alpha\beta}(\infty)\sigma_{\beta^{-1}}(1)
\sigma_{\alpha^{-1}}(0)\rangle\eqn\factor$$
Using the methods of conformal field theory, we write
$$\eqalign{&\langle\tau_{\beta}(z_1,\bar z_1)\tau'_{\alpha}(z_2,\bar z_2)
\sigma_{(\alpha\beta)^{-1}}(z_3,\bar z_3)\rangle=(-1)^{-\eta_\beta}
Y^{qu}_{\tau_{\beta}
\tau'_{\alpha}\sigma_{(\alpha\beta)^{-1}}}\cr &
(z_1-z_2)^{h_{\sigma_{(\alpha\beta)^{-1}}}-h_{\tau'_\alpha}-h_{\tau_\beta}}
(z_2-z_3)^{h_{\tau_\beta}-h_{\tau'_\alpha}-h_{\sigma_{(\alpha\beta)}^{-1}}}
(z_1-z_3)^{h_{\tau'_\alpha}-h_{\tau_\beta}-
h_{\sigma_{(\alpha\beta)}^{-1}}}\cr &
(\bar z_1-\bar z_2)^{\bar
h_{\sigma_{(\alpha\beta)^{-1}}}-\bar h_{\tau_\beta}-\bar h_{\tau'_\alpha}}
(\bar
z_2-\bar
z_3)^{\bar
h_{\tau_\beta}-\bar
h_{\tau'_\alpha}-\bar
h_{\sigma_{(\alpha\beta)^{-1}}}}(\bar
z_1-\bar z_3)^{\bar h_{\tau'_\alpha}-\bar h_{\tau_\beta}-
\bar h_{\sigma_{(\alpha\beta)^{-1}}}},}\eqn\confo$$ which implies that,
$$\langle\tau_{\beta}(\infty)\tau'_{\alpha}(x)
\sigma_{(\alpha\beta)^{-1}}(0)\rangle=(-1)^{-\eta_\beta}
Y_{\tau_{\beta}
\tau'_{\alpha}\sigma_{(\alpha\beta)^{-1}}}^{qu}
|x|^{2(h_{\sigma_\beta}-h_{\sigma_\alpha}-h_{{\sigma}_
{{(\alpha\beta)}^{-1}}})}x^{\eta_\alpha+\eta_\beta-1}.\eqn\addi$$
Then, inserting \addi\ in \factor, we obtain
$$\eqalign{\lim_{x,\bar x\rightarrow\infty}
\Big({\cal Z}^4_{excited}\Big)_{qu}=&
\lim_{x,\bar x\rightarrow\infty}(-1)^{-\eta_\beta}
|x|^{2(h_{\sigma_\beta}-h_{\sigma_\alpha}-h_{{\sigma}_
{{(\alpha\beta)}^{-1}}})}x^{\eta_\alpha+\eta_\beta-1}
\cr &Y^{qu}_{\tau_{\beta}
\tau'_{\alpha}\sigma_{{(\alpha\beta)}^{-1}}}
Y^{qu}_{\sigma_{\beta}
\sigma_{\alpha}\sigma_{{(\alpha\beta)}^{-1}}}}\eqn\may$$
where $Y^{qu}$ denotes the quantum contribution to the full Yukawa coupling.

Using \hell, we finally obtain
$$\Big({\cal Z}^4_{excited}\Big)_{qu}=
{1\over\sqrt{\eta_\beta(1-\eta_\alpha)}}(-1)^{-\eta_\beta}
x^{\eta_\alpha}(x-1)^{\eta_\beta-1}\Big((1-x)\partial_x
logI\Big){\cal Z}^4_{qu}.\eqn\help$$
The factorization of \help\ is now straightforward.
Using the fomulae of ref [\ten] for ${\cal Z}^4_{qu}$ and
the behaviour of the hypergeometic functions as $x\rightarrow \infty,$
we get
$$\eqalign{\lim_{x,\bar x\rightarrow \infty}& \Big({\cal
Z}^4_{excited}\Big)_{qu}=
(-1)^{-\eta_\beta}\lim_{x,\bar x\rightarrow \infty}
|x|^
{2(h_{\sigma_\beta}-h_{\sigma_\alpha}-h_{\sigma_
{(\alpha\beta)}^{-1}})}x^{\eta_\alpha+\eta_\beta-1}{\sqrt{(1-\eta_\alpha)\over\eta_\beta}}\cr &
\sqrt{{2V_\Lambda\over
ctg(\pi(1-\eta_\alpha))+ctg(\pi(1-\eta_\beta))}
}{\Gamma(\eta_\alpha)\Gamma(\eta_\beta)\over
\Gamma(\eta_\alpha+\eta_\beta-1)}
Y^{qu}_{\sigma_{\beta}
\sigma_{\alpha}\sigma_{{(\alpha\beta)}^{-1}}}}\eqn\maya$$
for $\eta_\beta>1-\eta_\alpha,$ where $\Lambda$ is the volume of the lattice
unit cell,
and
$$\eqalign{&\lim_{x,\bar x\rightarrow \infty} \Big({\cal
Z}^4_{excited}\Big)_{qu}=
(-1)^{-\eta_\beta}\lim_{x,\bar x\rightarrow \infty}
|x|^
{2(h_{\sigma_\beta}-h_{\sigma_\alpha}-h_{\sigma_
{(\alpha\beta)}^{-1}})}x^{\eta_\alpha+\eta_\beta-1}\cr &
{\sqrt{\eta_\beta\over(1-\eta_\alpha)}}\sqrt{{2V_\Lambda\over
ctg(\pi\eta_\alpha)+ctg(\pi\eta_\beta)}}
{\Gamma(1-\eta_\alpha)\Gamma(1-\eta_\beta)\over
\Gamma(1-\eta_\alpha-\eta_\beta)}Y^{qu}_{\sigma_{\beta}
\sigma_{\alpha}\sigma_{{(\alpha\beta)}^{-1}}}}\eqn\mayar$$
for $\eta_\beta<1-\eta_\alpha$.
{}From the equations \may, \maya\ and \mayar\ one can read off the
quantum part of the
Yukawa coupling \thpf\ as
$$
Y^{qu}_{\tau_{\beta}
\tau'_{\alpha}\sigma_{{(\alpha\beta)}^{-1}}}=
{\sqrt{(1-\eta_\alpha)\over\eta_\beta}}\sqrt{{2V_\Lambda\over
ctg(\pi(1-\eta_\alpha))+ctg(\pi(1-\eta_\beta))}
}{\Gamma(\eta_\alpha)\Gamma(\eta_\beta)\over
\Gamma(\eta_\alpha+\eta_\beta-1)}
\eqn\vit$$
for $\eta_\beta>1-\eta_\alpha$,
and
$$Y^{qu}_{\tau_\beta\tau'_\alpha\sigma_{(\alpha\beta)}^{-1}}=
{\sqrt{\eta_\beta\over(1-\eta_\alpha)}}{\sqrt{2V_\Lambda\over
ctg(\pi\eta_\alpha)+ctg(\pi\eta_\beta)}}
{\Gamma(1-\eta_\alpha)\Gamma(1-\eta_\beta)\over
\Gamma(1-\eta_\alpha-\eta_\beta)}\eqn\pinhead$$
for $\eta_\beta<1-\eta_\alpha$.

The classical part of
$Y_{\tau_{\beta}\tau'_{\alpha}\sigma_{{\alpha\beta}^{-1}}}$, which is
basically that for
$Y_{\sigma_{\beta}\sigma_{\alpha}\sigma_{{\alpha\beta}^{-1}}}$,
can be calculated by factorizing the classical part
$\sum_{X_{cl}}e^{-S (X_{cl})}$ in \plan\
or alternatively it could be calculated directly using the same techniques as
in the four-point function
and the fact that only one of the fields $\partial_z X_{cl}$ and
$\partial_z \bar X_{cl}$ has a non-zero solution.
The classical part of all Yukawa couplings for
${\bf Z}_M\times {\bf Z}_N$  Coxeter orbifolds has been calculated
in [\twelve].

\chapter {Excited Twisted Sector Yukawa couplings for ${\bf Z}_N$ Orbifolds}
In this section we consider the calculations of  the possible Yukawa
couplings involving excited twisted sectors in ${\bf Z}_N$ orbifolds.
Here it is found that no such couplings exist in the ${\bf Z}_3$,
${\bf Z}_4,$ ${\bf Z}_7,$
${\bf Z}_6$-${\bf I}$ and ${\bf Z}_8$-${\bf I}$ models for massless states with
quark and
lepton quantum numbers. In the rest of the
${\bf Z}_N$ models, the allowed couplings found are among
the excited twist fields of the type
$\tau_\alpha$, $\tau'_\alpha$, $\hat\tau_\alpha$ and $\hat\tau'_\alpha,$ where
$\hat\tau_\alpha$ and $\hat\tau'_\alpha$ are the doubly excited twist
fields mentioned before eqn. \medi.
In addition to the couplings  appearing in ${\bf Z}_N\times {\bf Z}_M,$ two
new kinds are found. These
are
$$\eqalign{&\langle\hat\tau_\beta(z_1,\bar z_1)\hat\tau'_\alpha(z_2,\bar z_2)
\sigma_{(\alpha\beta)^{-1}}(z_3,\bar z_3)
\rangle,\cr
&\langle\tau_\beta(z_1,\bar z_2)\tau_\alpha(z_2,\bar z_2)
\hat\tau'_{(\alpha\beta)^{-1}}(z_3,\bar z_3)\rangle.}\eqn\nv$$
All possible Yukawa couplings among excited twist fields for
${\bf Z}_N$ orbifolds are tabulated
in the Appendix, using the notation $T_p$ to denote the $\theta^p$
twisted sector.

The three-point functions in \nv\ can be evaluated using the same
procedure outlined in the previous section.
The first one can be written as
$$\langle\hat\tau_\beta(z_1,\bar z_1\hat\tau'_\alpha(z_2,\bar z_2)
\sigma_{(\alpha\beta)^{-1}}(z_3,\bar z_3)
\rangle=Y^{qu}_{\hat\tau_{\beta}
\hat\tau'_{\alpha}\sigma_{{(\alpha\beta)}^{-1}}}
Y^{cl}_{\hat\tau_{\beta}
\hat\tau'_{\alpha}\sigma_{{(\alpha\beta)}^{-1}}}
,\eqn\no$$
where the moduli-dependent classical part contained in
$Y^{cl}_{\hat\tau_{\beta}
\hat\tau'_{\alpha}\sigma_{{(\alpha\beta)}^{-1}}}$
is the same as for ground twist field three-point function.
The quantum piece is obtainable from the knowledge
of the four point function,
$${\cal Z}^4_{excited}={\cal
N}\langle\sigma_{\alpha^{-1}}(z_1){\hat\tau}'_\alpha(z_2)
\sigma_{\beta^{-1}}(z_3)
\hat\tau_{\beta}(z_4)\rangle,\eqn\nora$$
where ${\cal N}={1/ 8(1-\eta_\alpha)\eta_\beta}$ is a normalization factor due
to the fields
$\hat\tau'_\alpha$ and $\hat\tau_{\beta}.$
With  the aid of the operator product expansions \local,
the four point function \nora\ can be expressed as
$$\eqalign{{\cal Z}^4_{excited}=&{\cal N}
\lim_{z,z'\rightarrow z_4\atop
w,w'\rightarrow z_2} (z-z_4)^{1-\eta_\beta}(z'-z_4)^{1-\eta_\beta}
(w-z_2)^{\eta_\alpha}(w'-z_2)^{\eta_\alpha}\cr
&\langle\partial_{z}{X}\partial_{z'}{X}\partial_{w}{\bar
X}\partial_{w'}{\bar X}\sigma_{\alpha^{-1}}(z_1,{\bar z}_1)
{\sigma_\alpha}(z_2, {\bar
z}_2){\sigma}_{\beta^{-1}}(z_3, {\bar z}_3) \sigma_{\beta}(z_4,
{\bar z}_4)\rangle\cr &
={\cal N}\lim_{z,z'\rightarrow z_4\atop
w,w'\rightarrow z_2} (z-z_4)^{1-\eta_\beta}(z'-z_4)^{1-\eta_\beta}
(w-z_2)^{\eta_\alpha}(w'-z_2)^{\eta_\alpha}\sum_{X_{cl}}e^{-S_{cl}}\cr
&
\Big({\langle\partial_{z}{X}_{qu}\partial_{z'}{X}_{qu}\partial_{w}{\bar X}_{qu}
\partial_{w'}{\bar
X}_{qu}\rangle}_{{4-twists}}+
\partial_{z}{X}_{cl}\partial_{z'}{X}_{cl}\partial_{w}{\bar X}_{cl}
\partial_{w'}{\bar X}_{cl}{\cal Z}_{qu}^4\Big),}\eqn\plant$$
where the notation used is that of the previous section.
In order to obtain the quantum piece of the Yukawa coupling
\no, we need to factorize
the quantum part of the first term in \plant\ as the second
term makes zero contribution
(following the discussion of the previous section).
This term can be expressed as
$$\eqalign{{\Big({\cal Z}_4^{excited}\Big)_{qu}}=
&\lim_{{z,z'\rightarrow z_4}\atop w,w'\rightarrow
z_2}{\cal N}(z-z_4)^{1-\eta_\beta}(z'-z_4)^{1-\eta_\beta}
(w-z_2)^{\eta_\alpha}
(w'-z_2)^{\eta_\alpha}\cr &\langle\partial_{z}{X}_{qu}\partial_{z'}{X}_{qu}
\partial_{w}{\bar X}_{qu}
\partial_{w'}{\bar
X}_{qu}\rangle_{4-twists}\cr &
=\lim_{{z,z'\rightarrow z_4}\atop w,w'\rightarrow
z_2}
4{\cal N}(z-z_4)^{1-\eta_\beta}(z'-z_4)^{1-\eta_\beta}
(w-z_2)^{\eta_\alpha}
(w'-z_2)^{\eta_\alpha}\cr &
H(z, w, z',w',z_i)
\langle\sigma_{\alpha^{-1}}(z_1,{\bar z}_1)
{\sigma_\alpha}(z_2, {\bar z}_2){\sigma}_{\beta^{-1}}(z_3, {\bar
z}_3) \sigma_{\beta}(z_4, {\bar z}_4)\rangle}\eqn\lib$$
where
$$H(z, w, z',w')=
{1\over4}{\langle\partial_{z}{X}_{qu}\partial_{w}{\bar X}
\partial_{z'}{X}_{qu}\partial_{w'}{\bar X}\rangle_{4-twists}\over\langle
\sigma_{\alpha^{-1}}(z_1, \bar z_1)\sigma_{\alpha}(z_2,\bar z_2)
\sigma_{\beta^{-1}}(z_3,\bar z_3)\sigma_{\beta}(z_4,\bar z_4)\rangle}.\eqn\re$$
The correlator \re\ is calculated using local and global monodromy
conditions and is given by
$$\eqalign{H(z, w, z',w',z_i)=&
(z-z_1)^{-\eta_\alpha}(z-z_2)^{\eta_\alpha-1}(z-z_3)^{-\eta_\beta}
(z-z_4)^{\eta_\beta-1}\cr
&(z'-z_1)^{-\eta_\alpha}(z'-z_2)^{\eta_\alpha-1}(z'-z_3)^{-\eta_\beta}
(z'-z_4)^{\eta_\beta-1}\cr
&(w-z_1)^{\eta_\alpha-1}(w-z_2)^{-\eta_\alpha}
(w-z_3)^{\eta_\beta-1}
(w-z_4)^{-\eta_\beta}\cr &
(w'-z_1)^{\eta_\alpha-1}(w'-z_2)^{-\eta_\alpha}
(w'-z_3)^{\eta_\beta-1}
(w'-z_4)^{-\eta_\beta}\cr
&\Big[\Big({P(z,w)\over (z-w)^2}+A(z_i)\Big)\Big({P(z',w')\over
(z'-w')^2}+A(z_i)\Big)\cr & +\Big({P(z,w')\over
(z-w')^2}+A(z_i)\Big)\Big({P(z',w)\over
(z'-w)^2}+A(z_i)\Big)\Big].}\eqn\brigit$$
Then, performing the various limits in \lib\ and setting $z_1=0$, $z_2=x$,
$z_3=1$ and $z_4=\infty$,
we obtain
$${\Big({\cal Z}_4^{excited}\Big)}_{qu}={1\over \eta_\beta(1-\eta_\alpha)}
(-1)^{-2\eta_\beta}
x^{2\eta_\alpha}(x-1)^{2\eta_\beta-2}
\Big((1-x)\partial_x log I\Big)^2 {\cal Z}_{qu}^4\eqn\adil$$
To factorize into
$Y^{qu}_{\hat\tau_{\beta}
\hat\tau'_{\alpha}\sigma_{{(\alpha\beta)}^{-1}}}$, we take the limit
$x\rightarrow \infty,$ in \adil. In this limit we have
$$\lim_{x,\bar x\rightarrow\infty}
{\Big(Z_4^{excited}\Big)}_{qu}=\lim_{x,\bar x\rightarrow
\infty}
\langle\hat\tau_\beta(\infty){\hat\tau}'_\alpha(x)
\sigma_{{(\alpha\beta)}^{-1}}(0)\rangle
\langle\sigma_{\alpha\beta}(\infty)\sigma_{\beta^{-1}}(1)
\sigma_{\alpha^{-1}}(0)\rangle\eqn\carol$$
Using the methods of conformal field theory, one can write
$$\eqalign{&\langle\hat\tau_{\beta}(z_1)\hat\tau'_{\alpha}(z_2)
\sigma_{(\alpha\beta)^{-1}}(z_3)\rangle=(-1)^{-2\eta_\beta}
Y^{qu}_{\hat\tau_{\beta}
\hat\tau'_{\alpha}\sigma_{(\alpha\beta)^{-1}}}\cr &
(z_1-z_2)^{h_{\sigma_{(\alpha\beta)^{-1}}}-h_{\hat\tau'_\alpha}
-h_{\hat\tau_\beta}}
(z_2-z_3)^{h_{\hat\tau_\beta}-h_{\hat\tau'_\alpha}-h_{\sigma_{(\alpha\beta)}^{-1}}}
(z_1-z_3)^{h_{\hat\tau'_\alpha}-h_{\hat\tau_\beta}-
h_{\sigma_{(\alpha\beta)}^{-1}}}\cr &
(\bar z_1-\bar z_2)^{\bar
h_{\sigma_{(\alpha\beta)^{-1}}}-
\bar h_{\hat\tau_\beta}-\bar h_{\hat\tau'_\alpha}}
(\bar
z_2-\bar
z_3)^{\bar
h_{\hat\tau_\beta}-\bar
h_{\hat\tau'_\alpha}-\bar
h_{\sigma_{(\alpha\beta)^{-1}}}}(\bar
z_1-\bar z_3)^{\bar h_{\hat\tau'_\alpha}-\bar h_{\hat\tau_\beta}-
\bar h_{\sigma_{(\alpha\beta)^{-1}}}},}\eqn\confo$$ from which we obtain
$$\langle\hat\tau_{\beta}(\infty)\hat\tau'_{\alpha}(x)
\sigma_{(\alpha\beta)^{-1}}(0)\rangle=(-1)^{-2\eta_\beta}
Y_{\hat\tau_{\beta}
\hat\tau'_{\alpha}\sigma_{(\alpha\beta)^{-1}}}
|x|^{2(h_{\sigma_\beta}-h_{\sigma_\alpha}-h_{{\sigma}_
{{(\alpha\beta)}^{-1}}})}x^{2\eta_\alpha+2\eta_\beta-2}\eqn\ad$$
Then, substituting \ad\ in \carol, we obtain
$$\eqalign{&\lim_{x,\bar x\rightarrow\infty}
{\Big({\cal Z}_4^{excited}\Big)}_{qu}=\cr &
\lim_{x,\bar x\rightarrow\infty} (-1)^{-2\eta_\beta}
|x|^{2(h_{\sigma_\beta}-h_{\sigma_\alpha}-h_{{\sigma}_
{{(\alpha\beta)}^{-1}}})}x^{2\eta_\alpha+2\eta_\beta-2}
Y^{qu}_{\hat\tau_{\beta}
\hat\tau'_{\alpha}\sigma_{{(\alpha\beta)}^{-1}}}
Y^{qu}_{\sigma_{\beta}
\sigma_{\alpha}\sigma_{{(\alpha\beta)}^{-1}}}.}\eqn\hor$$
{}From the asymptotic nature of the hypergeometic functions as $x\rightarrow
\infty$, we obtain from \adil\
$$\eqalign{\lim_{x,\bar x\rightarrow \infty}&
\Big({\cal Z}^4_{excited}\Big)_{qu}= (-1)^{-2\eta_\beta}
\lim_{x,\bar x\rightarrow \infty}
|x|^
{2(h_{\sigma_\beta}-h_{\sigma_\alpha}-h_{\sigma_
{(\alpha\beta)}^{-1}})}x^{2\eta_\alpha+2\eta_\beta-2}{{(1-\eta_\alpha)\over\eta_\beta}}\cr
& \sqrt{{2V_\Lambda\over
ctg(\pi(1-\eta_\alpha))+ctg(\pi(1-\eta_\beta))}
}{\Gamma(\eta_\alpha)\Gamma(\eta_\beta)\over
\Gamma(\eta_\alpha+\eta_\beta-1)}
Y^{qu}_{\sigma_{\beta}
\sigma_{\alpha}\sigma_{{(\alpha\beta)}^{-1}}}}\eqn\horr$$
for $\eta_\beta>1-\eta_\alpha,$
and
$$\eqalign{\lim_{x,\bar x\rightarrow \infty}
\Big({\cal Z}^4_{excited}\Big)_{qu}=&(-1)^{-2\eta_\beta}
\lim_{x,\bar x\rightarrow \infty}
|x|^
{2(h_{\sigma_\beta}-h_{\sigma_\alpha}-h_{\sigma_
{(\alpha\beta)}^{-1}})}x^{2\eta_\alpha+2\eta_\beta-2}
{{\eta_\beta\over(1-\eta_\alpha)}}\cr &
\sqrt{{2V_\Lambda\over
ctg(\pi\eta_\alpha)+ctg(\pi\eta_\beta)}}
{\Gamma(1-\eta_\alpha)\Gamma(1-\eta_\beta)\over
\Gamma(1-\eta_\alpha-\eta_\beta)}Y^{qu}_{\sigma_{\beta}
\sigma_{\alpha}\sigma_{{(\alpha\beta)}^{-1}}}}\eqn\horrr$$
for $\eta_\beta<1-\eta_\alpha$.
{}From \hor, \horr\ and \horrr, one can read off the quantum part of the Yukawa
coupling
\lib\
$$
Y^{qu}_{\hat\tau_{\beta}
{\hat\tau}'_{\alpha}\sigma_{{(\alpha\beta)}^{-1}}}=
{{(1-\eta_\alpha)\over\eta_\beta}}\sqrt{{2V_\Lambda\over
ctg(\pi(1-\eta_\alpha))+ctg(\pi(1-\eta_\beta))}
}{\Gamma(\eta_\alpha)\Gamma(\eta_\beta)\over
\Gamma(\eta_\alpha+\eta_\beta-1)}\eqn\ha
$$
for $\eta_\beta>1-\eta_\alpha$,
and
$$Y^{qu}_{\hat\tau_{\beta}
{\hat\tau}'_{\alpha}\sigma_{{(\alpha\beta)}^{-1}}}=
{{\eta_\beta\over(1-\eta_\alpha)}}\sqrt{{2V_\Lambda\over
ctg(\pi\eta_\alpha)+ctg(\pi\eta_\beta)}}
{\Gamma(1-\eta_\alpha)\Gamma(1-\eta_\beta)\over
\Gamma(1-\eta_\alpha-\eta_\beta)}\eqn\haa$$
for $\eta_\beta<1-\eta_\alpha$.

Finally the second Yukawa coupling in \nv\ can be expressed as
$$\langle\tau_\beta(z_1,\bar z_1)\tau_\alpha(z_2,\bar z_2)
\hat\tau'_{(\alpha\beta)^{-1}}(z_3,\bar z_3)
\rangle=Y^{qu}_{\tau_{\beta}
\tau_{\alpha}\hat\tau'_{{(\alpha\beta)}^{-1}}}
Y^{cl}_{\tau_{\beta}
\tau_{\alpha}\hat\tau'_{{(\alpha\beta)}^{-1}}}.\eqn\paul$$
Here the moduli-dependent classical part contained in
$Y^{cl}_{\tau_{\beta}
\tau_{\alpha}\hat\tau'_{{(\alpha\beta)}^{-1}}}$
is the same as for ground state twist field three-point functions.
The quantum piece $
Y^{qu}_{\tau_{\beta}
\tau_{\alpha}\hat\tau'_{{(\alpha\beta)}^{-1}}}$
is obtainable through the knowledge
of the four point function
$$\eqalign{{\Big({\cal Z}_4^{excited}\Big)}_{qu}=
&{\cal M}\lim_{{{z\rightarrow z_1\atop w\rightarrow
z_2}\atop z'\rightarrow z_3}\atop w'\rightarrow z_4}
(z-z_1)^{\eta_\alpha}(w-z_2)^{1-\eta_\beta}(z'-z_3)^{\eta_\beta}
(w'-z_4)^{\eta_\beta}\cr &\langle\partial_{z}{X}_{qu}\partial_{w}{\bar X}
\partial_{z'}{X}_{qu}\partial_{w'}{\bar X}\rangle_{4-twists}
\cr &
=4{\cal M}\lim_{{{z\rightarrow z_1\atop w\rightarrow
z_2}\atop z'\rightarrow z_3}\atop w'\rightarrow z_4}
(z-z_1)^{\eta_\alpha}(w-z_2)^{1-\eta_\beta}(z'-z_3)^{\eta_\beta}
(w'-z_4)^{\eta_\beta}\cr & H(z, w, z',w',z_i)
\langle\sigma_{\alpha^{-1}}(z_1,{\bar z}_1)
{\sigma_\alpha}(z_2, {\bar z}_2){\sigma}_{\beta^{-1}}(z_3, {\bar
z}_3) \sigma_{\beta}(z_4, {\bar z}_4)\rangle}\eqn\lin$$
where ${\cal M}={1/4(1-\eta_\alpha)(1-\eta_\beta)}$ is a normalization factor.
By performing the various limits in \lin\ and setting $z_1=0$, $z_2=x$, $z_3=1$
and $z_4=\infty$,
we obtain
$$\eqalign{{\Big({\cal Z}_4^{excited}\Big)}_{qu}=&4{\cal
M}(-1)^{-\eta_\alpha-1}(x)^{2\eta_\alpha}(x-1)^{\eta_\beta-1}(1-x)^{\eta_\alpha-1}\cr
&\Big[
\Big(1-{\eta_\alpha\over2}-{\eta_\beta\over2})-
{\eta_\alpha-\eta_\beta\over2}
+(x-1)\partial_x log
I\Big)^2\cr
&+\Big({1\over2}(1-\eta_\alpha)(1-{2\over x})+
({1-\eta_\beta\over2})-{\eta_\alpha-\eta_\beta\over2}+(x-1)\partial_x
logI\Big)\cr &\Big(({1-\eta_\beta\over2})(1-{2\over
x})+{1\over2}(1-\eta_\alpha)-{\eta_\alpha-\eta_\beta\over2}+(x-1)\partial_x
logI\Big)\Big]{\cal Z}^4_{qu}}\eqn\moon$$

To factorize into $Y^{qu}_{\tau_{\beta}
\tau_{\alpha}\hat\tau'_{{(\alpha\beta)}^{-1}}}$, we take the limit
$x\rightarrow \infty,$ in \moon. In this limit,
using the methods of conformal field theory leads to
$$\lim_{x,\bar x\rightarrow \infty}
{\Big({\cal Z}_4^{excited}\Big)}_{qu}=
\lim_{x,\bar x\rightarrow \infty}
(x)^{3\eta_\alpha+\eta_\beta-2}
|x|^{2(h_{\sigma_\beta}-h_{\sigma_\alpha}-h_{{\sigma}_
{{(\alpha\beta)}^{-1}}})}
\Big(Y^{qu}_{\tau_{\beta}
\tau_{\alpha}\hat\tau'_{{(\alpha\beta)}^{-1}}}\Big)^2.\eqn\sad$$

{}From the behaviour of the hypergeometic functions as $x\rightarrow
\infty$, we find from \moon\
$$\eqalign{\lim_{x,\bar x\rightarrow \infty}
{\Big({\cal Z}_4^{excited}\Big)}_{qu}
&=
\lim_{x,\bar x\rightarrow \infty}
|x|^
{2(h_{\sigma_\beta}-h_{\sigma_\alpha}-h_{\sigma_
{(\alpha\beta)}^{-1}})}x^{3\eta_\alpha+\eta_\beta-2}\cr &
{(1-\eta_\beta-\eta_\alpha)^2\over
(1-\eta_\alpha)(1-\eta_\beta)}{4V_\Lambda\over
ctg(\pi\eta_\alpha)+ctg(\pi\eta_\beta)}
{\Gamma^2(1-\eta_\alpha)\Gamma^2(1-\eta_\beta)\over
\Gamma^2(1-(\eta_\alpha+\eta_\beta))}}\eqn\eva$$
for $\eta_\beta<1-\eta_\alpha$, and vanishing otherwise.

Then equations \sad\ and \eva\ lead to
$$Y^{qu}_{\tau_{\beta}
\tau_{\alpha}\hat\tau'_{{(\alpha\beta)}^{-1}}}=
2\sqrt{{V_{\Lambda}\over ctg(\pi\eta_\alpha)+ctg(\pi\eta_\beta)}}
{\Gamma(1-\eta_\alpha)\Gamma(1-\eta_\beta)\over
\Gamma(1-(\eta_\alpha+\eta_\beta))}
{(1-\eta_\beta-\eta_\alpha)\over
\sqrt{(1-\eta_\alpha)(1-\eta_\beta)}}.\eqn\fin$$
The classical part of the Yukawa couplings for certain Coxeter
${\bf Z}_N$ orbifolds has been calculated in [\eight].

In summary, we have calculated the Yukawa couplings
involving excited twist fields in
all ${\bf Z}_N$ and ${\bf Z}_M\times {\bf Z}_N$  Coxeter orbifolds.
We have limited
ourselves to those of phenomenological interest, i.e., excited twist
fields with gauge quantum numbers of quarks, leptons and Higgses of the
standard model.
However, it is found that only the quantum part of these couplings is
modified relative to the Yukawa couplings of unexcited twist fields. The
classical part coming from
worldsheet instantons is the same as before. Therefore our analysis is
valid for all ${\bf Z}_N$ and ${\bf Z}_M\times {\bf Z}_N$  orbifolds and not
only the Coxeter
type.
Our results may be of significance in obtaining the detailed pattern of quark
and lepton masses

\centerline{ACKNOWLEDGEMENTS}
This work was supported in part by S.E.R.C.
\vfill\eject

\chapter {APPENDIX}
Possible excited three-point functions ${\Big({\cal
Z}_i^3\Big)}_{excited}$ for the ${\bf Z}_N$
and ${\bf Z}_M\times {\bf Z}_N$ orbifolds. The point group elements $\alpha$
associated with the various fields are represented by the subscript
$\eta_\alpha.$
\vskip 0.5cm
\input tables
\hfill\break
\centerline{${\bf Z}_6$-${\bf II}$}
\vskip 0.5cm
\begintable
Yukawa couplings|Possible excited three-point functions\cr
$T_1T_1T_4$|$\langle\tau^1_{1/6}\sigma^1_{1/6}\tau'^1_{4/6}\rangle$,
$\langle\sigma^1_{1/6}\tau^1_{1/6}\tau'^1_{4/6}\rangle$
\endtable
\vskip 0.5cm
\hfill\break
\centerline {${\bf Z}_8$-${\bf II}$}
\vskip 0.5cm
\begintable
Yukawa couplings|Possible excited three-point functions\cr
$T_1T_1T_6$|$\langle\tau^1_{1/8}\sigma^1_{1/8}\tau'^1_{6/8}\rangle$,
$\langle\sigma^1_{1/8}\tau^1_{1/8}\tau'^1_{6/8}\rangle$
\endtable
\vskip 0.5cm
\hfill\break
\centerline{${\bf Z}_{12}$-${\bf I}$}
\vskip 0.5cm
\begintable
Yukawa couplings|Possible excited three-point functions\cr
$T_2T_3T_7$|$\langle\tau^3_{2/12}\tau'^3_{9/12}\sigma^3_{1/12}\rangle$,
$\langle\sigma^3_{2/12}\tau'^3_{9/12}\tau^3_{1/12}\rangle$\cr
$T_1T_2T_9$|$\langle\sigma^1_{1/12}\tau^1_{2/12}\tau'^1_{9/12}\rangle$,
$\langle\tau^1_{1/12}\sigma^1_{2/12}\tau'^1_{9/12}\rangle$
\endtable
\vskip 0.5cm
\hfill\break
\vfill\eject
\centerline{${\bf Z}_{12}$-${\bf II}$}
\vskip 0.5cm
\begintable
Yukawa couplings|Possible excited three-point functions\cr
$T_1T_3T_8$|$\langle\tau^1_{1/12}\sigma^1_{3/12}\tau'^1_{8/12}\rangle$,
$\langle\sigma^1_{1/12}\tau^1_{3/12}\tau'^1_{8/12}\rangle$\cr
$T_1T_1T_{10}$|$\langle\tau^1_{1/12}\sigma^1_{1/12}\tau'^1_{10/12}\rangle$,
$\langle\sigma^1_{1/12}\tau^1_{1/12}\tau'^1_{10/12}\rangle$,
$\langle{\hat\tau}^1_{1/12}\sigma^1_{1/12}{\hat\tau}'^1_{10/12}\rangle$,\nr
|
$\langle\sigma^1_{1/12}{\hat\tau}^1_{1/12}{\hat\tau}'^1_{10/12}\rangle$,
$\langle\tau^1_{1/12}\tau^1_{1/12}{\hat\tau}'^1_{10/12}\rangle$\cr
$T_2T_5T_{5}$|
$\langle\tau'^2_{10/12}\tau^2_{1/12}\sigma^2_{1/12}\rangle$,
$\langle\tau'^2_{10/12}\sigma^2_{1/12}\tau^2_{1/12}\rangle$,
$\langle{\hat\tau}'^2_{10/12}{\hat\tau}^2_{1/12}\sigma^2_{1/12}\rangle$,\nr |
$\langle{\hat\tau}'^2_{10/12}\sigma^2_{1/12}{\hat\tau}^2_{1/12}\rangle$,
$\langle{\hat\tau}'^2_{10/12}\tau^2_{1/12}\tau^2_{1/12}\rangle$,
\endtable
\vfill\eject
\hfill\break
\centerline{${\bf Z}_3\times {\bf Z}_6$}
\vskip 0.5cm
\begintable
Yukawa Coupling|Possible excited three-point
functions\cr
$T_{01}T_{14}T_{21}$|
$\langle\sigma^2_{1/6}{\tau}'{}^2_{2/3}
{\tau}^2_{1/6}\rangle$,\nr
|
$\langle{\tau}^2_{1/6}{\tau}'{}^2_{2/3}\sigma^2_{1/6}\rangle$
\cr $T_{02}T_{13}T_{21}$|
$\langle{\tau}'{}^3_{2/3}{\tau}^3_{1/6}
\sigma^3_{1/6}\rangle$,\nr |
$\langle{\tau}'{}^3_{2/3}\sigma^3_{1/6}{\tau}^3_{1/6}\rangle$\cr
$T_{04}T_{11}T_{21}$|$\langle{\tau}'{}^2_{2/3}
{\tau}^2_{1/6} \sigma^2_{1/6}\rangle$,
\nr | $\langle{\tau}'{}^2_{2/3}\sigma^2_{1/6}
{\tau}^2_{1/6}\rangle$\cr
$T_{05}T_{10}T_{21}$|
$\langle{\tau}^3_{1/6}{\tau}'{}^3_{2/3}
\sigma^3_{1/6}\rangle$,
\nr | $\langle\sigma^3_{1/6}{\tau}'{}^3_{2/3}
{\tau}^3_{1/6}\rangle$
\cr $T_{10}T_{13}T_{13}$|
$\langle{\tau}'{}^3_{2/3}{\tau}^3_{1/6}
\sigma^3_{1/6}\rangle$,\nr
| $\langle{\tau}'{}^3_{2/3}\sigma^3_{1/6}
{\tau}^3_{1/6}\rangle$
\cr  $T_{11}T_{11}T_{14}$|
$\langle{\tau}^2_{1/6}\sigma^2_{1/6}
{\tau}'{}^2_{2/3}\rangle$,\nr
| $\langle\sigma^2_{1/6}{\tau}^2_{1/6}
{\tau}'{}^2_{2/3}\rangle$
\endtable
\vfill\eject
\centerline{${\bf Z}_2\times {\bf Z}'_6$}
\vskip 0.5cm
\begintable
Yukawa Coupling|Possible excited three-point
functions\cr
$T_{01}T_{11}T_{14}$|
$\langle{\tau}^1_{1/6}{\tau'}^1_{4/6}\sigma^1_{1/6}\rangle$,
$\langle{\tau}^2_{1/6}\sigma^2_{1/6}{\tau'}^2_{4/6}\rangle$,
$\langle{\tau'}^3_{4/6}{\tau}^3_{1/6}\sigma^3_{1/6}\rangle$,
$\langle\sigma^2_{1/6}{\tau}^2_{1/6}{\tau'}^2_{4/6}\rangle$,
$\langle{\tau'}^3_{4/6}\sigma^3_{1/6}{\tau}^3_{1/6}\rangle$,\nr |
$\langle{\tau'}^3_{4/6}{\tau}^3_{1/6}\sigma^3_{1/6}\rangle\langle
\sigma^2_{1/6}{\tau}^2_{1/6}{\tau'}^2_{4/6}\rangle$,\nr |
$\langle{\tau}^1_{1/6}{\tau'}^1_{4/6}\sigma^1_{1/6}\rangle
\langle{\tau}^2_{1/6}\sigma^2_{1/6}{\tau'}^2_{4/6}\rangle$,
$\langle\sigma^1_{1/6}{\tau'}^1_{4/6}{\tau}^1_{1/6}\rangle$,
$\langle{\tau'}^3_{4/6}\sigma^3_{1/6}{\tau}^3_{1/6}\rangle
\langle\sigma^1_{1/6}{\tau'}^1_{4/6}{\tau}^1_{1/6}
\rangle$
\endtable
\vskip 0.5cm
\hfill\break
\centerline{${\bf Z}_2\times {\bf Z}_6$}
\vskip 0.5cm
\begintable
Yukawa Coupling|Possible excited three-point
functions\cr
$T_{04}T_{11}T_{11}$|
$\langle{\tau'}^2_{4/6}{\tau}^2_{1/6}\sigma^2_{1/6}\rangle$,
$\langle{\tau'}^2_{4/6}\sigma^2_{1/6}{\tau}^2_{1/6}\rangle$
\endtable
\vskip 0.5cm
\hfill\break
\centerline{${\bf Z}_6\times {\bf Z}_6$}
\vskip 0.5cm
\begintable
Coupling|Possible excited three-point functions\cr
$T_{01}T_{14}T_{51}$|
$\langle\tau^2_{1/6}\tau'^2_{4/6}\sigma^2_{1/6}\rangle$,
$\langle\sigma^2_{1/6}\tau'^2_{4/6}\tau^2_{1/6}\rangle$\cr
$T_{01}T_{24}T_{41}$| $\langle\tau^2_{1/6}\tau'^2_{4/6}\sigma^2_{1/6}\rangle$,
$\langle\sigma^2_{1/6}\tau'^2_{4/6}\tau^2_{1/6}\rangle$\cr
$T_{02}T_{14}T_{50}$|
$\langle\tau'^3_{4/6}\tau^3_{1/6}\sigma^3_{1/6}\rangle$,
$\langle\tau'^3_{4/6}\sigma^3_{1/6}\tau^3_{1/6}\rangle$\cr
$T_{02}T_{23}T_{41}$|$\langle\tau'^3_{4/6}\tau^3_{1/6}\sigma^3_{1/6}\rangle$,
$\langle\tau'^3_{4/6}\sigma^3_{1/6}\tau^3_{1/6}\rangle$\cr
$T_{02}T_{32}T_{32}$|$\langle\tau'^3_{4/6}\tau^3_{1/6}\sigma^3_{1/6}\rangle$,
$\langle\tau'^3_{4/6}\sigma^3_{1/6}\tau^3_{1/6}\rangle$\cr
$T_{04}T_{11}T_{51}$|
$\langle\tau'^2_{4/6}\tau^2_{1/6}\sigma^2_{1/6}\rangle$,
$\langle\tau'^2_{4/6}\sigma^2_{1/6}\tau^2_{1/6}\rangle$
\endtable
continued on next page
\vfill\eject
\begintable
$T_{04}T_{21}T_{41}$|$\langle\tau'^2_{4/6}\sigma^2_{1/6}\tau^2_{1/6}\rangle$,
$\langle\tau'^2_{4/6}\tau^2_{1/6}\sigma^2_{1/6}\rangle$ \cr
$T_{05}T_{11}T_{50}$|$\langle\tau^3_{1/6}
\tau'^3_{4/6}\sigma^3_{1/6}\rangle$,$\langle\sigma^3_{1/6}\tau'^3_{4/6}
\tau^3_{1/6}\rangle$\cr
$T_{05}T_{20}T_{41}$|$\langle\tau^3_{1/6}\tau'^3_{4/6}\sigma^3_{1/6}\rangle$,
$\langle\sigma^3_{1/6}\tau'^3_{4/6}\tau^3_{1/6}\rangle$\cr
$T_{10}T_{14}T_{42}$|
$\langle\tau^1_{1/6}\sigma^1_{1/6}\tau'^1_{4/6}\rangle$,
$\langle\sigma^1_{1/6}\tau^1_{1/6}\tau'^1_{4/6}\rangle$\cr
$T_{10}T_{15}T_{41}$|
$\langle\tau^1_{1/6}\sigma^1_{1/6}\tau'^1_{4/6}\rangle$,
$\langle\sigma^1_{1/6}\tau^1_{1/6}\tau'^1_{4/6}\rangle$\cr
$T_{20}T_{23}T_{23}$|$\langle\tau'^3_{4/6}\tau^3_{1/6}\sigma^3_{1/6}\rangle$,
$\langle{\tau}'^3_{4/6}\sigma^3_{1/6}\tau^3_{1/6}\rangle$\cr
$T_{11}T_{13}T_{42}$|$\langle\tau^1_{1/6}\sigma^1_{1/6}\tau'^1_{4/6}\rangle$,
$\langle\sigma^1_{1/6}\tau^1_{1/6}\tau'^1_{4/6}\rangle$\cr
$T_{11}T_{14}T_{41}$|
$\langle\tau^1_{1/6}\sigma^1_{1/6}\tau'^1_{4/6}\rangle$,
$\langle\tau^2_{1/6}\tau'^2_{4/6}\sigma^2_{1/6}\rangle$,
$\langle\sigma^1_{1/6}\tau^1_{1/6}\tau'^1_{4/6}\rangle$,
$\langle\sigma^2_{1/6}\tau'^2_{4/6}\tau^2_{1/6}\rangle$,
$\langle\tau'^3_{4/6}\sigma^3_{1/6}\tau^3_{1/6}\rangle$,\nr
|$\langle\tau'^3_{4/6}\tau^3_{1/6}\sigma^3_{1/6}\rangle$,\nr |
$\langle\tau^1_{1/6}\sigma^1_{1/6}\tau'^1_{4/6}\rangle\langle
\tau^2_{1/6}\tau'^2_{4/6}\sigma^2_{1/6}\rangle$,
$\langle\tau'^3_{4/6}\sigma^3_{1/6}\tau^3_{1/6}\rangle\langle
\sigma^2_{1/6}\tau'^2_{4/6}\tau^2_{1/6}\rangle
$,
$\langle\tau'^3_{4/6}\tau^3_{1/6}\sigma^3_{1/6}\rangle
\langle\sigma^1_{1/6}\tau^1_{1/6}\tau'^1_{4/6}\rangle$
\cr
$T_{11}T_{15}T_{40}$|$\langle\tau^1_{1/6}
\sigma^1_{1/6}\tau'^1_{4/6}\rangle$,$\langle\sigma^1_{1/6}\tau^1_{1/6}
\tau'^1_{4/6}\rangle$\cr
$T_{11}T_{23}T_{32}$|$\langle\tau'^3_{4/6}
\tau^3_{1/6}\sigma^3_{1/6}\rangle$,
$\langle\tau'^3_{4/6}\sigma^3_{1/6}\tau^3_{1/6}\rangle$\cr
$T_{11}T_{24}T_{31}$|$\langle\tau^2_{1/6}\tau'^2_{4/6}\sigma^2_{1/6}\rangle$,
$\langle\sigma^2_{1/6}\tau'^2_{4/6}\tau^2_{1/6}\rangle$\cr
$T_{12}T_{12}T_{42}$|$\langle\tau^1_{1/6}\sigma^1_{1/6}\tau'^1_{4/6}\rangle$,
$\langle\sigma^1_{1/6}\tau^1_{1/6}\tau'^1_{4/6}\rangle$\cr
$T_{12}T_{13}T_{41}$|$\langle\tau^1_{1/6}\sigma^1_{1/6}\tau'^1_{4/6}\rangle$,
$\langle\sigma^1_{1/6}\tau^1_{1/6}\tau'^1_{4/6}\rangle$\cr
$T_{12}T_{14}T_{40}$|$\langle\tau^1_{1/6}\sigma^1_{1/6}\tau'^1_{4/6}\rangle$,
$\langle\sigma^1_{1/6}\tau^1_{1/6}\tau'^1_{4/6}\rangle$\cr
$T_{13}T_{13}T_{40}$|$\langle\tau^1_{1/6}\sigma^1_{1/6}\tau'^1_{4/6}\rangle$,
$\langle\sigma^1_{1/6}\tau^1_{1/6}\tau'^1_{4/6}\rangle$ \cr
$T_{14}T_{20}T_{32}$|$\langle\tau^3_{1/6}\tau'^3_{4/6}\sigma^3_{1/6}\rangle$,
$\langle\sigma^3_{1/6}\tau'^3_{4/6}\tau^3_{1/6}\rangle$ \cr
$T_{14}T_{21}T_{31}$|$\langle\tau'^2_{4/6}\sigma^2_{1/6}\tau^2_{1/6}\rangle$,
$\langle\tau'^2_{4/6}\tau^2_{1/6}\sigma^2_{1/6}\rangle$\cr
$T_{21}T_{21}T_{24}$|$\langle\tau^2_{1/6}\sigma^2_{1/6}\tau'^2_{4/6}\rangle$,
$\langle\sigma^2_{1/6}\tau^2_{1/6}\tau'^2_{4/6}\rangle$
\endtable
\refout
\bye